\documentclass[12pt]{article}

\usepackage{graphicx,amssymb}
\usepackage{hyperref}
\usepackage{amsmath}

\usepackage{amssymb}
\newcounter{comment}

{\refstepcounter{comment}%
\begin{quote}
\ttfamily\small$\blacksquare$ \textbf{\underline{Comment} $\sharp$\thecomment:}}%
{\end{quote}}

{
\begin{quote}
\ttfamily\small$\blacktriangleright$ \textbf{\underline{Reply} $\sharp$\thecomment:}}%
{\end{quote}}

\newcommand{\GeV}{{\rm GeV}}
\font\cmss=cmss12 
\def\1{\hbox{{1}\kern-.25em\hbox{l}}}
\def\bfZ{\relax{\hbox{\cmss Z\kern-.4em Z}}}

\begin{document}

\begin{titlepage}

\centerline{\large \bf Deeply virtual Compton scattering  beyond
next-to-leading order: } \vspace{2mm} \centerline{\large \bf the
flavor singlet case}

\vspace{10mm}

\centerline{\bf K.~Kumeri{\v c}ki$^{a,b}$,  D.~M\"uller$^{c}$,
K.~Passek-Kumeri{\v c}ki$^{a,d}$, and A.~Sch\"afer$^{a}$}

\vspace{8mm}

\centerline{\it $^a$Institut f\"ur Theoretische Physik,
Universit\"at Regensburg} \centerline{\it D-93040 Regensburg,
Germany}

\vspace{8mm}

\centerline{\it $^b$Department of Physics,
Faculty of Science, University of Zagreb}
\centerline{\it P.O.B. 331, HR-10002 Zagreb, Croatia}

\vspace{8mm}

\centerline{\it $^c$Department of Physics and Astronomy, Arizona
State University} \centerline{\it Tempe, AZ 85287-1504, USA}

\vspace{8mm}
\centerline{\it $^d$ Theoretical Physics Division,
Rudjer Bo{\v s}kovi{\'c} Institute}
\centerline{\it P.O.Box 180, HR-10002 Zagreb, Croatia}

\vspace{3mm}

\vspace{5mm}

\centerline{\bf Abstract}
          \noindent We study radiative corrections to deeply
          virtual Compton scattering in the kinematics of HERA
          collider experiments to next--to--leading and
          next--to--next--to--leading order. In the latter case
          the radiative corrections are evaluated in a special
          scheme that allows us to employ the predictive power of
          conformal symmetry. As observed before, the size of
          next--to--leading order corrections strongly depends on
          the gluonic input, as gluons start to contribute at this
          order. Beyond next--to--leading order we find, in
          contrast, that the corrections for an input scale
          of few $\mbox{GeV}^2$
          are small enough to justify the
          uses of perturbation theory. For $\xi \gtrsim 5\cdot
          10^{-3}$  the modification of the scale dependence is
          also small. However, with decreasing $\xi$ it becomes
          moderate or even large, in particular for the phase.

\vspace{0.5cm}

\noindent

\vspace*{12mm}


\end{titlepage}

\section{Introduction}

Deeply virtual Compton scattering (DVCS) is considered as the
theoretically cleanest process to investigate generalized parton
distributions (GPDs) \cite{MueRobGeyDitHor94,Ji96,Rad96}. These
distributions are a hybrid of parton densities, form-factors, and
distribution amplitudes and might be represented in terms of
light--cone wave functions \cite{DieFelJakKro00, BroDieHwa00, BroChaHarMukVar06}.
They are rather intricate functions depending on the longitudinal momentum
fractions in the $s$-- and $t$--channels, the momentum transfer
squared, and the resolution scale. On the other hand they are
phenomenologically very attractive, since they allow to combine
information from different experiments in an optimal manner and
since they encode non--perturbative information that cannot be
extracted from either inclusive or elastic measurements alone.
Their second moments provide e.g. the total angular momentum of
partons in the nucleon \cite{Ji96} and the gravitational form
factor of the nucleon. Moreover, a  specific partial
Fourier-transformation of GPDs is phenomenologically very
interesting, as it provides functions which have a probabilistic
interpretation. In the impact parameter space they can be viewed
as  parton densities in dependence of the longitudinal momentum
fraction and the transverse distance from the proton center
\cite{Bur00,Bur02,BelMue02}. The knowledge of transverse parton
distribution does not only add substantially to our understanding
of hadron structure but is also relevant for the prediction of
cross sections in dependence of the impact parameter. For
proton--proton scattering this has been especially emphasized with
respect to LHC physics in Ref.\ \cite{FraStrWei05}.

On one hand GPDs are thus a new window to study  non--perturbative
QCD and it has been already impressively demonstrated that they
are experimentally accessible via  the DVCS process
\cite{Airetal01,Steetal01,Clas06,Adletal01,Chekanov:2003ya,Aktas:2005ty}.
On the other hand for several theoretical and experimental reasons
the extraction of GPDs from measurements remains quite challenging
because typically one is sensitive only to convolutions containing
GPDs and one has to disentangle different contributions. The
analysis simplifies substantially at large energies, where both photon
\cite{Adletal01,Chekanov:2003ya,Aktas:2005ty}  and
vector--meson  leptoproduction, e.g., Refs.\ \cite{Bre98,Adl99,Che02},  have
been measured by H1 and ZEUS. In HERA kinematics the
photon--proton interaction starts to be flavor blind and so one
mainly accesses flavor singlet GPDs. Moreover, spin flip effects
are suppressed, too, and only one set of GPDs is relevant, namely,
the proton helicity conserved and  parton helicity averaged GPDs
$H(x,\xi,t,{\cal Q}^2)$. In this paper we concentrate on these
singlet NNLO corrections, as the non-singlet  case has already
been studied in \cite{Mue05a}. As usual, such analysis is only
possible after adopting some parametrization for the
dependence of GPDs on the $s$-- and $t$--channel momentum fraction. We
believe that realistic models can be most easily constructed by
means of the partial wave decomposition of GPDs and amplitudes
\cite{MueSch05,ManKirSch05,KirManSch05a}, where the dominant
contributions arise from the leading Regge trajectories
\cite{MueSch05,Mue06,SzcLon06}. Support for this conjecture arises
also from lattice calculations \cite{Hagetal03, Gocetal03,
Hagetal04,Hag04a,Gocetal05}.

In this letter we study radiative corrections to DVCS at and beyond
next--to--leading (NLO) order. This investigation is partially
motivated by the fact that within a certain class of GPD models
perturbative corrections at this order were reported to be rather
large \cite{BelMueNieSch99,FreMcD01b} and so one should worry
about the justification of the perturbative QCD approach.
Recently, the radiative corrections in the flavor \emph{non--singlet}
case have been studied and it has been concluded that the relative
radiative corrections are moderate at NLO and become smaller at
next--to--next--to--leading order (NNLO).
Hence, these findings support the perturbative formalism in
this sector. Considering the \emph{singlet} case at hand,
we recall that the leading order (LO) contribution is
given by the quark handbag diagram and that at next--to--leading
order the gluon distribution appears as a new entry. It is
known from deeply inelastic scattering (DIS) that the gluonic contribution
in the small $x$--region is much larger than that of the sea quarks. Hence,
the size of the NLO corrections depends in particular on the
gluonic GPD and the appearance of large NLO corrections does not
necessarily mean that perturbation theory fails.

To clarify the situation, we employ here conformal symmetry to
obtain the next--to--next--to--leading order corrections of the
DVCS amplitude. In Sect.\ \ref{Sec-ConApp} we present, after a
short introduction to the conformal approach, the analytic result
for the DVCS amplitude in NNLO. In Sect.\ \ref{Sec-NumAna},
relying on the pomeron pole as the dominant contribution at small
momentum fraction, we numerically evaluate radiative corrections
up to NNLO for the kinematics of HERA collider experiments. This
analysis includes a comparison of the standard predictions in NLO
with those of the conformal approach. We finally give our
conclusions in Sect.\ \ref{Sec-Con}.

\section{The DVCS amplitude in NLO and NNLO}
\label{Sec-ConApp}

The DVCS amplitude is defined in terms of the hadronic tensor
\begin{eqnarray}
\label{Def-HadTen} T_{\mu\nu} (q, P_1, P_2) = \frac{i}{e^2} \int
d^4{}x\, e^{i x\cdot q} \langle P_2, S_2 | T j_\mu (x/2) j_\nu (-x/2) |
P_1, S_1 \rangle,
\end{eqnarray}
where $q = (q_1 + q_2)/2$ ($\mu$  and $q_2$ refers to the outgoing
real photon). To leading power in ${\cal Q}^2=-q_1^2$
(leading twist) and LO in the QCD coupling constant
the hadronic tensor (\ref{Def-HadTen}) is
evaluated from the hand--bag
diagram. In terms of the kinematical variables
$P=P_1+P_2$ and $\Delta=P_2-P_1$, the result can be written as
\begin{eqnarray}
        \label{decom-T} T_{\mu\nu} (q,P,\Delta) = - \left(\widetilde{g}_{\mu\nu} -\frac{\widetilde{P_\mu q_\nu}}{P\cdot q}-
\frac{\widetilde{P_\nu q_\mu}}{P\cdot q}\right)\frac{q_\sigma V^\sigma}{P\cdot q} - i
\widetilde{\epsilon}_{\mu \nu q P} \frac{q_\sigma
A^\sigma}{(P\cdot q)^2} + \cdots .
\end{eqnarray}
where the tilde--symbol denotes contraction
$\widetilde{X}_{\mu\nu} \equiv {\cal P}_{\mu\rho}\, X^{\rho\sigma}\,
{\cal P}_{\sigma\nu}$
with projectors
\begin{equation}
{\cal P}^{\alpha \beta} = g^{\alpha\beta} -\frac{q_1^\alpha
q^\beta_2}{q_1\cdot q_2} \;,
\end{equation}
to ensure current conservation \cite{BelMueNieSch00}. The ellipsis
indicate terms that are finally power suppressed in the DVCS amplitude or are determined by the
gluon transversity GPD, which is suppressed by $\alpha_s/\pi$ and is not considered here.
Note that to leading twist accuracy the parenthesis in (\ref{decom-T}) can be replaced by
$\widetilde{g}_{\mu\nu}$.
In the parity even sector the vector
\begin{eqnarray}
\label{dec-FF-V} V^{\sigma} = \overline{U} (P_2, S_2) \left( {\cal
H} \gamma^\sigma + {\cal E} \frac{i\sigma^{\sigma\rho}
\Delta_\rho}{2M} \right) U (P_1, S_1) + \cdots\, ,
\end{eqnarray}
is decomposed into the target helicity  conserving Compton form
factor (CFF) ${\cal H}$ and the helicity flip one ${\cal E}$.
Analogously, the axial-vector
\begin{eqnarray}
\label{dec-FF-A} A^{\sigma} = \overline{U} (P_2,S_2) \left(
\widetilde{\cal H} \gamma^\sigma\gamma_5 + \widetilde{\cal E}
\frac{\Delta^\sigma \gamma_5}{2M} \right) U (P_1,S_1) + \cdots ,
\end{eqnarray}
is parametrized in terms of $\widetilde{\cal
H}$ and $\widetilde{\cal E}$, where again higher twist
contributions are neglected. The normalization of the spinors is
$\overline{U} (p, S) \gamma^\sigma U (p, S) = 2 p^\sigma$. We also introduce
the scaling variables
\begin{eqnarray}
\xi = \frac{Q^2}{P\cdot q}\,,\qquad \eta = -\frac{\Delta\cdot
q}{P\cdot q}\,,
\end{eqnarray}
where $Q^2=-q^2$. In DVCS kinematics and twist--two accuracy we
have $\xi=\eta$, while ${\cal  Q}^2= 2 Q^2$.

Before we proceed, let us decompose the CFFs, denoted by the set
${\cal F} = \{{\cal H},{\cal E},\widetilde {\cal H},\widetilde
{\cal E}\}$, in flavor non--singlet (NS) and singlet (S) ones:
\begin{eqnarray}
{\cal F}  = Q_{\rm NS}^2\,  {^{\rm NS}\!{\cal F}} +  Q_{\rm S}^2\,
{^{\rm S}\! {\cal F}}\,, \qquad  {^{\rm S}\!{\cal F}} ={^{\Sigma}\!{\cal F}} +
{^{\rm G}\!{\cal F}}\, ,
\end{eqnarray}
where the singlet piece contains the quark flavor singlet ${^{\Sigma}\! {\cal
F}}$ and gluon ${^{\rm G}\!{\cal F}}$ CFFs. The charge factors
$Q_i^2$ with $i=\{{\rm NS},{\rm S}\}$ are given as linear
combination of squared quark charges, e.g., the singlet
one is given by the average of the squared charges for $n_f$
active quarks:
\begin{equation}
Q_{\rm S}^2 = \frac{1}{n_f} \sum_{i=u,d,\cdots} Q_i^2\,.
\end{equation}
In the momentum fraction representation the  Compton form factors
are represented as convolution of the coefficient function with
the corresponding GPD. In the singlet sector in which quark
(${^{\Sigma}\!{\cal O}}$) and gluon (${^{\rm G}\!{\cal O}}$) operators
mix under renormalization, we might introduce the vector notation:
\begin{eqnarray}
\label{Def-CFF} {^{\rm S}\! {\cal F}}(\xi,\Delta^2,{\cal Q}^2) =
\int_{-1}^{1}\! \frac{dx}{\xi}\ \mbox{\boldmath $C$}(x/\xi,{\cal
Q}^2/\mu^2,\alpha_s(\mu)|\xi) \mbox{\boldmath $ F$}(x,\eta=\xi,
\Delta^2,\mu^2)\,.
\end{eqnarray}
Here the column vector
\begin{eqnarray}
\mbox{\boldmath $F$} = \left({ {^{\Sigma}\!F }\atop {^{\rm G}\! F}
}\right)\,, \quad  {F} = \{{H},{E},\widetilde {H},\widetilde
{E}\}
\end{eqnarray}
contains the GPDs, and the row one, defined as $\mbox{\boldmath $C$}=({^{\Sigma}\!C}, \;
(1/\xi) {^{\rm G}\! C})$, consists of the hard scattering part that to LO
accuracy reads
\begin{eqnarray}
\label{Def-Cquark} \frac{1}{\xi} \mbox{\boldmath $C$}(x/\xi,{\cal
Q}^2/\mu^2,\alpha_s(\mu)|\xi)
=\left(\frac{1}{\xi-x-i\epsilon},0\right) + {\cal O}(\alpha_s)\,.
\end{eqnarray}
We remark that the $\xi$ dependence in ${^\Sigma\!C}$ and ${^{\rm G}\!C}$
enters only via the ratio $x/\xi$. Note also that the
$u$--channel contribution in the quark entry (\ref{Def-Cquark})
has been reabsorbed into the symmetrized quark singlet
distribution
\begin{eqnarray}
{^\Sigma\!F}(x,\eta,\Delta^2,\mu^2) = \sum_{q=u,d,\cdots}
\left[{^q\!F}(x,\eta,\Delta^2,\mu^2)\mp
{^q\!F}(-x,\eta,\Delta^2,\mu^2)\right]\,.
\end{eqnarray}
Here the second term in the square brackets with $-(+)$--sign for
$H,\, E$ ($\widetilde H,\, \widetilde E$)-type GPDs is for
$x>\eta$ related to the s--channel exchange of an anti--quark.
The gluon GPDs have
definite symmetry property under the exchange of $x\to -x$:
${^{\rm G}\! H}$ and ${^{\rm G}\! E}$ are even, while
${^{\rm  G}\! \widetilde{H}}$ and ${^{\rm G}\! \widetilde{E}}$ are odd.

The convolution formula (\ref{Def-CFF}) has already at LO the
disadvantage that it contains a  singularity at the cross--over
point between the central region ($-\eta\leq x\leq \eta$) and the
outer region ($\eta\leq x\leq 1$), i.e., for $x=\xi=\eta$. Its
treatment is defined by the $i\epsilon$ prescription, coming from
the Feynman propagator. The GPD is considered smooth at this
point, but will generally not be holomorphic \cite{Rad97}. The fact that
both regions are dual to each other, up to a so--called $D$-term contribution \cite{PolWei99},
makes the numerical treatment even more complicated.
This motivated our development of a more suitable formalism in \cite{MueSch05}.
The factorization scale $\mu$ in the
GPDs is ambiguous and at LO order this induces the main
uncertainty. Beyond LO this factorization scale dependence will be
cancelled in the considered order of perturbation theory.
The NLO corrections to the coefficient functions \cite{BelMue97a} and to
the evolution kernels \cite{BelMueFre99,BelMue99c,BelFreMue99}
were predicted from conformal constraints, where the rotation to
the standard $\overline{\rm MS}$ scheme has been taken into
account. Note that the conformal symmetry in $\overline{\rm MS}$ scheme is broken
and that the predicted results coincide 
with the diagrammatic evaluation
\cite{ManPilSteVanWei97,JiOsb97,JiOsb98}. To this order and in
this scheme a numerical code has been made accessible that includes
evolution, see, e.g., \cite{FreMcD01b}. As already mentioned
above, it was found that the perturbative corrections to
NLO can be quite large.

At present it seems hardly possible to study perturbative
corrections beyond NLO accuracy in the standard scheme, since the
diagrammatical evaluation would require enormous effort.
Fortunately, we can employ conformal symmetry to relate the
perturbative corrections at NNLO to those for DIS
\cite{ZijNee92,ZijNee94,vanNeerven:2000uj}, where the NNLO
corrections in the vector case has been completed by the
substantial effort of Vogt, Moch and Vermaseren
\cite{VogMocVer04}. From these calculations we get the
normalization of the Wilson coefficients and anomalous dimensions.
The conformal predictions arise from the application of the
conformal operator product expansion and are valid as long as the
twist--two operators behave covariantly under conformal
transformation \cite{FerGriGat71,Mue97a,BraKorMue03}. This is
certainly true at tree level and  it also can be  ensured for
vanishing $\beta$--function in any order of perturbation theory
within a special renormalization scheme \cite{Mue97a}.  To make
contact with the conformal  OPE, we expand the hard--scattering
amplitude in terms of Gegenbauer polynomials with indices $3/2$
and  $5/2$ for quarks and gluons, respectively, and introduce the
conformal GPD moments, which formally leads to
\begin{eqnarray}
\label{Exp-CFFs}
 {^{\rm S}\! {\cal F}}(\xi,\Delta^2,{\cal Q}^2)  = 2 \sum_{j=0}^\infty \xi^{-j-1}  \mbox{\boldmath $C$}_{j}({\cal
Q}^2/\mu^2,\alpha_s(\mu))\; \mbox{\boldmath $F$}_{j}(\xi,\Delta^2,\mu^2).
\end{eqnarray}
The expansion coefficients $\mbox{\boldmath $C$}_{j}$
can be calculated by the projection:
\begin{multline}
\label{Def-HarSca2ConMom}
 \mbox{\boldmath $C$}_{j}
({\cal Q}^2/\mu^2,\alpha_s(\mu))
= \frac{
2^{j+1}\Gamma(j+5/2)}{\Gamma(3/2) \Gamma(j+4)} \\
\times \frac{1}{2}
\int_{-1}^1\! dx\;\mbox{\boldmath $C$}(x,{\cal Q}^2/\mu^2,\alpha_s(\mu)|\xi=1)
\left(
\begin{array}{cc}
 (j+3)[1-x^2] C_j^{3/2} & 0 \\
0 & 3[1-x^2]^2 C_{j-1}^{5/2}
\end{array}
\right)\!\left(x\right)\,.
\end{multline}
Note that we have here rescaled the integration variable with respect
to $\xi$ and that the integral runs only over the rescaled central
region.
The conformal moments of the singlet GPDs are defined as
\begin{eqnarray}
\label{Fj}
\mbox{\boldmath $F$}_{j}(\eta,\Delta^2,\mu^2) = \frac{
\Gamma(3/2)\Gamma(j+1)}{2^{j}  \Gamma(j+3/2)}  \frac{1}{2}\int_{-1}^1\!
dx\; \eta^{j-1} \left(
\begin{array}{cc}
\eta\, C_j^{3/2} & 0 \\
0 & (3/j)\,  C_{j-1}^{5/2}
\end{array}
\right)\!\!\left(\frac{x}{\eta}\right)
 \mbox{\boldmath $F$}(x,\eta,\Delta^2,\mu^2)\,.
\end{eqnarray}
Here $j$ is an odd (even) non-negative integer for the  (axial--)vector case.

In the forward kinematics ($\Delta\to 0$), our conventions are
such that the helicity conserved GPDs coincide with the flavor
singlet quark distribution and with $x$ times the gluon
distribution. Hence, for the moments we have agreement with the
common Mellin--moments of parton densities, e.g., for the helicity
averaged GPD:
\begin{eqnarray}
\lim_{\eta\to 0}\mbox{\boldmath $H$}(x,\eta) =
\left(
\begin{array}{c}
\Sigma \\
x G
\end{array}
\right)(x)\,,
\quad
\mbox{\boldmath $q$}_j\equiv
\lim_{\eta\to 0}\mbox{\boldmath $H$}_{j}(\eta)
  = \int_{0}^1\!dx\; x^j
\left(
\begin{array}{c}
\Sigma \\
G
\end{array}
\right)(x) \;.
\end{eqnarray}

Unfortunately, the series (\ref{Exp-CFFs}) does
not converge for DVCS kinematics, in particular not in the outer
region, and one has to resum the OPE \cite{MueSch05,Mue05a} or,
equivalently, one can use a dispersion relation \cite{Che97,KumMueKumPasSch06}.
The result for $^{\rm S}\mathcal{H}$  in terms of a
Mellin--Barnes integral reads
\begin{eqnarray}
\label{Res-ImReCFF} {^{\rm S}\!{\cal H}}(\xi,\Delta^2,{\cal Q}^2)
&\!\!\!=\!\!\!& \frac{1}{2i}\int_{c-i \infty}^{c+ i \infty}\!
dj\,\xi^{-j-1} \left[i +
\tan
\left(\frac{\pi j}{2}\right) \right] \mbox{\boldmath
$C$}_{j}({\cal Q}^2/\mu^2,\alpha_s(\mu)) \mbox{\boldmath $H$}_{j}
(\xi,\Delta^2,\mu^2)
 \,.
\end{eqnarray}
In the following we write the  perturbative expansion as
\begin{eqnarray}
\label{Res-WilCoe-Exp-CS-SI}
\lefteqn{\mbox{\boldmath $C$}_{j}^{}({\cal Q}^2/\mu^2,{\cal Q}^2/\mu_{r}^2,\alpha_s(\mu_r))}
\nonumber \\ &=&
 \frac{2^{j+1} \Gamma(j+5/2)}{\Gamma(3/2)\Gamma(j+3)}
\left[{\mbox{\boldmath $C$}_{j}^{(0)}}  +
\frac{\alpha_s(\mu_r)}{2\pi} \mbox{\boldmath $C$}_j^{ (1)}({\cal
Q}^2/\mu^2) + \frac{\alpha^2_s(\mu_r)}{(2\pi)^2} \mbox{\boldmath
$C$}_j^{(2)}({\cal Q}^2/\mu^2,{\cal Q}^2/\mu_{r}^2) + {\cal O}(\alpha_s^3) \right],\,
\qquad
\end{eqnarray}
where corresponding to our conventions, the LO  Wilson coefficients are normalized as
\begin{eqnarray}
{\mbox{\boldmath $C$}_{j}^{(0)}} 
= (1\;,\;0) \,,
\end{eqnarray}
and here we choose to distinguish the renormalization ($\mu_r$) and factorization
($\mu$) scales.

Let us first give here the DVCS NLO corrections in the $\overline{\rm MS}$ scheme.
We restrict ourselves to the analysis of the kinematically dominant
contribution, i.e., $^{\rm S}{\cal H}$, and so we provide here only the results for the vector case.
The conformal moments (\ref{Def-HarSca2ConMom}) can be obtained from Refs.\
\cite{BelMue97a,ManPilSteVanWei97,JiOsb97,JiOsb98}.
Using the representation of Ref.\ \cite{BelMueNieSch99},
the integrals which are needed are evaluated in a
straightforward manner  for integer conformal spin, see Appendix C of Ref.\ \cite{MelMuePas02} for
the quark entries. The analytic continuation to complex $j$ leads to
\begin{eqnarray}
^{\Sigma}\!C_j^{(1)}({\cal Q}/\mu^2)&\!\!\!=\!\!\!& C_F \left[ 2 S^2_{1}(1 +
j)- \frac{9}{2}  + \frac{5-4S_{1}(j+1)}{2(j + 1)(j + 2)} +
\frac{1}{(j+1)^2(j+2)^2}\right] + \frac{^{\Sigma\Sigma}\!\gamma_j^{(0)}}{2}
\ln\frac{\mu^2}{{\cal Q}^2} \, ,
\label{Res-WilCoe-MS-NLO-V}
\\
\label{Res-WilCoe-MS-NLO-Vg} ^{\rm G}\!C_j^{(1)}({\cal
Q}/\mu^2)&\!\!\!=\!\!\!& -2 n_f T_F\frac{(4 + 3j + j^2)
\left[S_{1}(j)+S_{1}(j+2)\right]  +2 + 3j + j^2}{
                                       ( 1 + j)( 2 + j)( 3 + j) }
+ \frac{^{\Sigma {\rm G}}\!\gamma_j^{(0)}}{2} \ln\frac{\mu^2}{{\cal Q}^2} \, ,
\end{eqnarray}
where $C_F=4/3$ and
$T_F=1/2$. The entries of the anomalous dimension matrix read at LO:
\begin{eqnarray}
{^{\Sigma\Sigma}\!\gamma}_{j}^{(0)} &\!\!\!=&\!\!\! - C_F \left( 3 + \frac{2}{(
j + 1 )( j + 2 )} - 4 S_{1}(j + 1) \right)
\\
{^{\Sigma{\rm G}}\!\gamma}_{j}^{(0)} &\!\!\!=&\!\!\! -4n_f T_F\frac{4
+ 3\,j + j^2 }{( j + 1 )( j + 2 )( j + 3)}\,,
\\
{^{{\rm G}\Sigma}\!\gamma}_{j}^{(0)} &\!\!\!=&\!\!\! -2C_F\frac{4 +
3\,j + j^2 }{j( j + 1 )( j + 2 )}\,,
\\
\label{Def-LO-AnoDim-GG-V} {^{\rm GG}\!\gamma}_{j}^{(0)}
&\!\!\!=&\!\!\! - C_A \left(-\frac{4}{( j + 1 )( j + 2
)}+\frac{12}{j( j + 3)} - 4S_1( j + 1 )  \right)+ \beta_0\,,
\end{eqnarray}
where $\beta_0 = 2 n_f/3 - 11 C_A/3$, $C_A=3$. In the
$\overline{\rm MS}$ scheme also the complete anomalous dimension
matrix is known to two-loop accuracy \cite{Mue94,BelMue98c}.
However, the conformal moments will mix with each other and
the solution of the evolution equation has so far not been given in terms
of a Mellin--Barnes integral.

The advantage of the conformal symmetry is that it predicts the
Wilson coefficients. However, the symmetry is only valid in a
special conformal scheme ($\overline{\rm CS}$  \footnote{The
treatment of the terms proportional to $\beta$, that break conformal
symmetry, is of course ambiguous. We employ here the so-called
$\overline{\rm CS}$ scheme in which the running of the coupling is
implemented in the form of the conformal operator product
expansion (COPE) that is valid for a
hypothetical fixed point. In particular, conformal moments are
multiplicatively renormalizable to NLO. For details see Refs.\
\cite{MelMuePas02,Mue05a}.}). In such a scheme the structure of the
Wilson coefficients up to NNLO is
\begin{eqnarray}
\label{Res-WilCoe-CS-NLO} \mbox{\boldmath $C$}_j^{(1)}({\cal
Q}^2/\mu^2) &\!\!\! =\!\!\!  & \mbox{\boldmath ${c}$}_j^{(1)}+
\frac{s^{(1)}_j({\cal Q}^2/\mu^2)}{2} \; \mbox{\boldmath
${c}$}_j^{(0)} \mbox{\boldmath ${\gamma}$}_j^{(0)}\,,
\\
\label{Res-WilCoe-CS-NNLO} \mbox{\boldmath $C$}_j^{(2)}({\cal
Q}^2/\mu^2,{\cal Q}^2/\mu_{r}^2) &\!\!\! =\!\!\!  & \mbox{\boldmath ${c}$}_j^{(2)} +
\frac{s^{(1)}_j({\cal Q}^2/\mu^2)}{2} \left[ \mbox{\boldmath
${c}$}_j^{(0)}\mbox{\boldmath ${\gamma}$}_j^{(1)} +
\mbox{\boldmath ${c}$}_j^{(1)} \mbox{\boldmath
${\gamma}$}_j^{(0)}\right] + \frac{s^{(2)}_j({\cal Q}^2/\mu^2)}{8}
\; \mbox{\boldmath ${c}$}_j^{(0)} \left(\mbox{\boldmath
${\gamma}$}_j^{(0)}\right)^2 \qquad
\\
&& +\frac{\beta_0}{2}
 \left[ \mbox{\boldmath $
C$}_j^{(1)}({\cal Q}^2/\mu^2)\ln\frac{{\cal Q}^2}{\mu_r^2} +
\frac{1}{4} \mbox{\boldmath ${c}$}_j^{(0)} \mbox{\boldmath
${\gamma}$}_j^{(0)} \ln^2\frac{{\cal Q}^2}{\mu^2} \right]\,,
 \nonumber
\end{eqnarray}
where the so-called shift coefficients $s_j^{(i)}({\cal Q}^2/\mu^2)$ can be
expressed in terms of harmonic sums $S_{p}(n) = \sum_{k=1}^n
1/k^p$ as
\begin{eqnarray}
\label{eq:s12Q2}
s_j^{(1)}({\cal Q}^2/\mu^2)&= &
  S_1(j+3/2)-S_1(j+2) +2\ln(2)-\ln\frac{{\cal Q}^2}{\mu^2}\,, \quad
 \\
s_j^{(2)}({\cal Q}^2/\mu^2)&=& \left(s_j^{(1)}({\cal
Q}^2/\mu^2)\right)^2
   -S_2(j+3/2)+ S_2(j+2) \,,
\end{eqnarray}
and $\mbox{\boldmath ${ c}$}_j^{(i)}= ({^\Sigma\! c_j^{(i)}},{^{\rm G}\! c_j^{(i)}})$ are the Wilson coefficients known from DIS.
We have at LO $\mbox{\boldmath ${ c}$}_j^{(0)}=(1,0)$,
at NLO
\begin{eqnarray}
\label{eq:NScV1}
 {^\Sigma\!c}_j^{(1)} \!\!\!&=&\!\!\!
     C_F \left[S^2_{1}(1 + j) + \frac{3}{2} S_{1}(j + 2)
      - \frac{9}{2}  + \frac{5-2S_{1}(j)}{2(j + 1)(j + 2)}
      - S_{2}(j + 1)\right]\,,
      \\
{^{\rm G}\!c}_j^{(1)}\!\!\!&=&\!\!\!
- 2 n_f T_{F} \frac{(4 + 3j + j^2)  S_{1}(j)  +2 + 3j + j^2}{( 1 + j)( 2 +
j)( 3 + j) }\;, \label{Def-Coe-NLO-G-V}
\end{eqnarray}
and at NNLO they are given by the Mellin moments of the DIS partonic structure
functions \cite{ZijNee92,ZijNee94}. To simplify their evaluation, we take for
$\mbox{\boldmath ${ c}$}_j^{(2)}$ a fit, given in \cite{vanNeerven:2000uj},
rather than the exact expression.

The evolution of the singlet (integer) conformal moments in this
$\overline{\rm CS}$ scheme is  governed by
\begin{eqnarray}
\label{Def-RGE-1} \mu\frac{d}{d\mu}
\mbox{\boldmath $F$}_{j}(\xi, \Delta^2,\mu^2) &\!\!\!=\!\!\!& -\Bigg[ \frac{\alpha_s(\mu)}{2\pi}
\mbox{\boldmath $\gamma$}^{(0)}_j +
\frac{\alpha_s^2(\mu)}{(2\pi)^2} \mbox{\boldmath
$\gamma$}_j^{(1)}+ \frac{\alpha_s^3(\mu)}{(2\pi)^3}
\mbox{\boldmath $\gamma$}_j^{(2)} +{\cal O}(\alpha_s^4) \Bigg]
\mbox{\boldmath $F$}_j(\xi, \Delta^2,\mu^2)
\nonumber\\
&&\hspace{0.5cm} -\frac{\beta_0}{2}
\frac{\alpha_s^3(\mu)}{(2\pi)^3}\sum_{k=0}^{j-2}
\left[\Delta_{jk}^{\overline{{\rm CS}}}+{\cal O}(\alpha_s)
\right]\mbox{\boldmath $F$}_k(\xi, \Delta^2,\mu^2)\,,
\end{eqnarray}
where the mixing matrix $\mbox{\boldmath
$\Delta$}_{jk}^{\overline{{\rm CS}}}$ is not completely known. In the vector
case the anomalous dimensions are known to NNLO \cite{VogMocVer04}.  In absence
of the mixing term, the solution of the renormalization group equation
$\mbox{\boldmath $F$}_{j}(\xi, \Delta^2,\mu^2)
=
 \mbox{\boldmath ${\cal E}$}_j(\mu,\mu_0)
\mbox{\boldmath $F$}_{j}(\xi, \Delta^2,\mu_0^2)$
can be given using the path--ordered exponential evolution operator
\begin{eqnarray}
\label{Def-EvoOpe}
 \mbox{\boldmath ${\cal E}$}_j(\mu,\mu_0) &\!\!\! =\!\!\! &   {\cal P}
 \exp{\left\{-\int_{\mu_0}^{\mu} \frac{d\mu^\prime}{\mu^\prime}
\mbox{\boldmath ${\gamma}$}_j (\alpha_s(\mu^\prime))\right\}}\,.
\end{eqnarray}
In the numerical analysis we will only resum the leading
logarithms and expand the non--leading ones
\begin{eqnarray}
\label{Exp--EvoOpe}
\mbox{\boldmath ${\cal E}$}_j(\mu,\mu_0) =  \sum_{a,b=\pm}\left[
    \delta_{ab}\, {^{a}\!\mbox{\boldmath $P$}}_j +
      \frac{\alpha_s(\mu)}{2\pi}\, {^{ab}\!\!\mbox{\boldmath ${\cal A}$}}_j^{(1)}(\mu,\mu_0)
      + \frac{\alpha_s^2(\mu)}{(2\pi)^2}\,  {^{ab}\!\!\mbox{\boldmath ${\cal A}$}}_j^{(2)}(\mu,\mu_0)
+O(\alpha_s^3) \right]\left[ \frac{\alpha_s(\mu)}{\alpha_s(\mu_0)}
\right]^{-\frac{{^b\! \lambda}_j}{\beta_0}}  .\nonumber\\
\end{eqnarray}
Here the projectors on the $\{+,-\}$ modes are
\begin{eqnarray}
{^{\pm}\!\mbox{\boldmath $P$}}_j = \frac{\pm 1}{{^{+}\!
\lambda}_j-{^{-}\! \lambda}_j} \left(\mbox{\boldmath
$\gamma$}_j^{(0)}-{^{\mp}\! \lambda}_j \mbox{\boldmath
$1$}\right)\,,
\end{eqnarray}
where the eigenvalues of the LO anomalous dimension matrix are
\begin{eqnarray}
{^{\pm}\! \lambda}_j
 =\frac{1}{2}\left({^{\Sigma\Sigma}\!
\gamma}_j^{(0)} + {^{\rm GG}\! \gamma}^{(0)}_j\right) \mp
\frac{1}{2} \left({^{\Sigma\Sigma}\! \gamma}^{(0)}_j - {^{\rm GG}\!
\gamma}^{(0)}_j\right) \sqrt{1 + \frac{4{^{\Sigma{\rm G}}\!
\gamma}^{(0)}_j {^{{\rm G}\Sigma}\! \gamma}^{(0)}_j}{\left({^{\Sigma\Sigma}\!
\gamma}^{(0)}_j - {^{\rm GG}\! \gamma}^{(0)}_j\right)^2}} \,.
\end{eqnarray}
A straightforward calculation leads to the matrix valued
coefficients
\begin{eqnarray}
{^{ab}\!\!\mbox{\boldmath ${\cal A}$}}_j^{(1)}  &\!\!\!=\!\!\!&
{^{ab}\!R}_j(\mu,\mu_0|1) {^{a}\!\mbox{\boldmath $P$}}_j \left[
\frac{\beta_1}{2\beta_0}  \mbox{\boldmath $\gamma$}_j^{(0)}
-\mbox{\boldmath $\gamma$}_j^{(1)} \right] {^{b}\!\mbox{\boldmath
$P$}}_j
\\
{^{ab}\!\!\mbox{\boldmath ${\cal A}$}}_j^{(2)} &\!\!\!=\!\!\!&
\sum_{c=\pm} \frac{1}{\beta_0+{^{c}\! \lambda}_j-{^{b}\!
\lambda}_j}\!\!\left[{^{ab}\!R}_j(\mu,\mu_0|2)-{^{ac}\!R}_j(\mu,\mu_0|1)\!
\left(\frac{\alpha_s(\mu_0)}{\alpha_s(\mu)}\right)^{\frac{\beta_0+{^{c}\!
\lambda}_j-{^{b}\! \lambda}_j}{\beta_0}} \right]
{^{a}\!\mbox{\boldmath $P$}}_j \left[ \frac{\beta_1}{2\beta_0}
\mbox{\boldmath $\gamma$}_j^{(0)} -\mbox{\boldmath
$\gamma$}_j^{(1)} \right]
\nonumber\\
&&\!\!\!\!\times {^{c}\!\mbox{\boldmath $P$}}_j\left[
\frac{\beta_1}{2\beta_0}  \mbox{\boldmath $\gamma$}_j^{(0)}
-\mbox{\boldmath $\gamma$}_j^{(1)} \right] {^{b}\!\mbox{\boldmath
$P$}}_j - {^{ab}\!R}_j(\mu,\mu_0|2) {^{a}\!\mbox{\boldmath
$P$}}_j\left[\frac{\beta_1^2-\beta_2 \beta_0}{4\beta_0^2}
 \mbox{\boldmath $\gamma$}_j^{(0)} -
\frac{\beta_1}{2\beta_0} \mbox{\boldmath $\gamma$}_j^{(1)} +
 \mbox{\boldmath $\gamma$}_j^{(2)}\right]{^{b}\!\mbox{\boldmath $P$}}_j\,,
\end{eqnarray}
where the $\mu$ dependence is accumulated in the following
functions:
\begin{eqnarray}
{^{ab}\!R}_j(\mu,\mu_0|n)= \frac{1}{ n \beta_0+{^{a}\!
\lambda}_j-{^{b}\! \lambda}_j}\left[ 1-
\left(\frac{\alpha_s(\mu_0)}{\alpha_s(\mu)}\right)^{\frac{n
\beta_0+{^{a}\! \lambda}_j-{^{b}\! \lambda}_j}{\beta_0}} \right] \;.
\end{eqnarray}
The expansion coefficients of the $\beta$ function are defined as
\begin{eqnarray}
&&\!\!\!\!\! \frac{\beta}{g} = \frac{\alpha_s(\mu)}{4 \pi} \beta_0
+ \frac{\alpha_s^2(\mu)}{(4 \pi)^2} \beta_1
+\frac{\alpha_s^3(\mu)}{(4 \pi)^3} \beta_2+ O(\alpha_s^4),
\\
&&\!\!\!\!\! \beta_0 = \frac{2}{3} n_f -11,\quad \beta_1 =
\frac{38}{3} n_f - 102, \quad \beta_2 = -\frac{325}{54} n_f^2 +
\frac{5033}{18} n_f - \frac{2857}{2}\, . \nonumber
\end{eqnarray}
The expansion of the evolution operator (\ref{Exp--EvoOpe}) will then be consistently combined with the
Wilson--coefficients (\ref{Res-WilCoe-Exp-CS-SI}), see for instance Ref.\
\cite{MelMuePas02}.

\section{Numerical evaluation of radiative corrections}
\label{Sec-NumAna}

Here we study the perturbative corrections in the small--$\xi$ region.
We adopt a simple ansatz for the conformal moments that is inspired
by the dominance of the pomeron and by the assumption that ${\cal
O}(\xi^2)$ terms in the conformal polynomials are insignificant:
\begin{eqnarray}
\label{Ans-ConMom}
\mbox{\boldmath $H$}_j(\xi, \Delta^2, {\cal Q}^2) = \left(
\begin{array}{c}
N_{\rm sea} {^{\rm sea}\!F}(\Delta^2)\, {\rm B}(1+j-\alpha_{\rm sea}(\Delta^2),8)/{\rm B}(2-\alpha_{\rm sea}(0),8) \\
N_{\rm G} {^{\rm G}\!F}(\Delta^2)\, {\rm B}(1+j-\alpha_{\rm G}(\Delta^2),6)/{\rm B}(2-\alpha_{\rm G}(0),6)
\end{array}
\right) + \cdots,
\end{eqnarray}
where ${\rm B}(x,n) =\Gamma(x)\Gamma(n)/\Gamma(x+n)$, and
the ellipsis denotes the neglected ${\cal O}(\xi^2)$ terms, as
well as valence components whose contributions are also
small for small $\xi$. Here the normalization is ${ ^{\rm
sea}\!F}(\Delta^2=0)={^{\rm G}\!F}(\Delta^2=0)=1$ and
for $\alpha_{i}(\Delta^2)$ we use the ``effective''
pomeron trajectory $\alpha(\Delta^2)= \alpha(0) +
\alpha^\prime \Delta^2 $. We remind that in deeply
inelastic scattering the structure function $F_2 \sim
(1/x_{\rm Bj})^{\lambda(Q^2)}$ grows with increasing $Q^2$.
Here the exponent is related to the intercept of the Regge
trajectory $\lambda=\alpha(0)-1$. The values for
$\alpha(0)$ will be specified below. Although also the
slope $\alpha^\prime$ is scale dependent \cite{Mue06},
we choose here the standard value of the soft pomeron
$\alpha^\prime=0.25$. In the forward case  the moments
(\ref{Ans-ConMom}) arise from the  parton densities
\begin{eqnarray}
\Sigma = \frac{N_{\rm sea}}{{\rm B}(2-\alpha_{\rm sea}(0),8)}\, x^{-\alpha_\Sigma(0)}\, (1-x)^7 + u_{\rm v}(x)+ d_{\rm v}(x)\,,
&\: & G = \frac{N_G}{{\rm B}(2-\alpha_{\rm G}(0),6)}\, x^{-\alpha_{\rm G}(0)} \, (1-x)^5 \,, \nonumber \\
\end{eqnarray}
for which we have adopted a generic realistic parametrization.
Note that the valence component $u_{\rm v}+ d_{\rm v}$ is not taken into account in Eq.\ (\ref{Ans-ConMom}),
since it is a non--leading contribution for small $x$.
The normalization factors are related by the momentum sum rule
\begin{eqnarray}
\int_0^1\!dx\, x\left[\Sigma(x) + G(x) \right]=1\quad\Rightarrow \quad N_{\rm G} + N_{\rm sea} +
\int_{0}^1\!dx\;x \left[u_{\rm v}+ d_{\rm v}\right](x) =1\,.
\end{eqnarray}
In the asymptotic limit $\cal Q\rightarrow \infty$, the evolution equation
tells us that
$ N_{\rm G} = 4 C_F/(4 C_F+n_f)$ i.e. that more than $50\%$ of the longitudinal proton
momentum is carried by gluons. At experimentally accessible large scales the
gluons already carry about $40\,\%$ of the momentum. For the momentum of the
valence quarks we choose the generic value $1/3$ and so $N_{\rm sea}= 2/3-
N_{\rm G}$.

\begin{figure}[t]
\includegraphics[clip,scale=0.72]{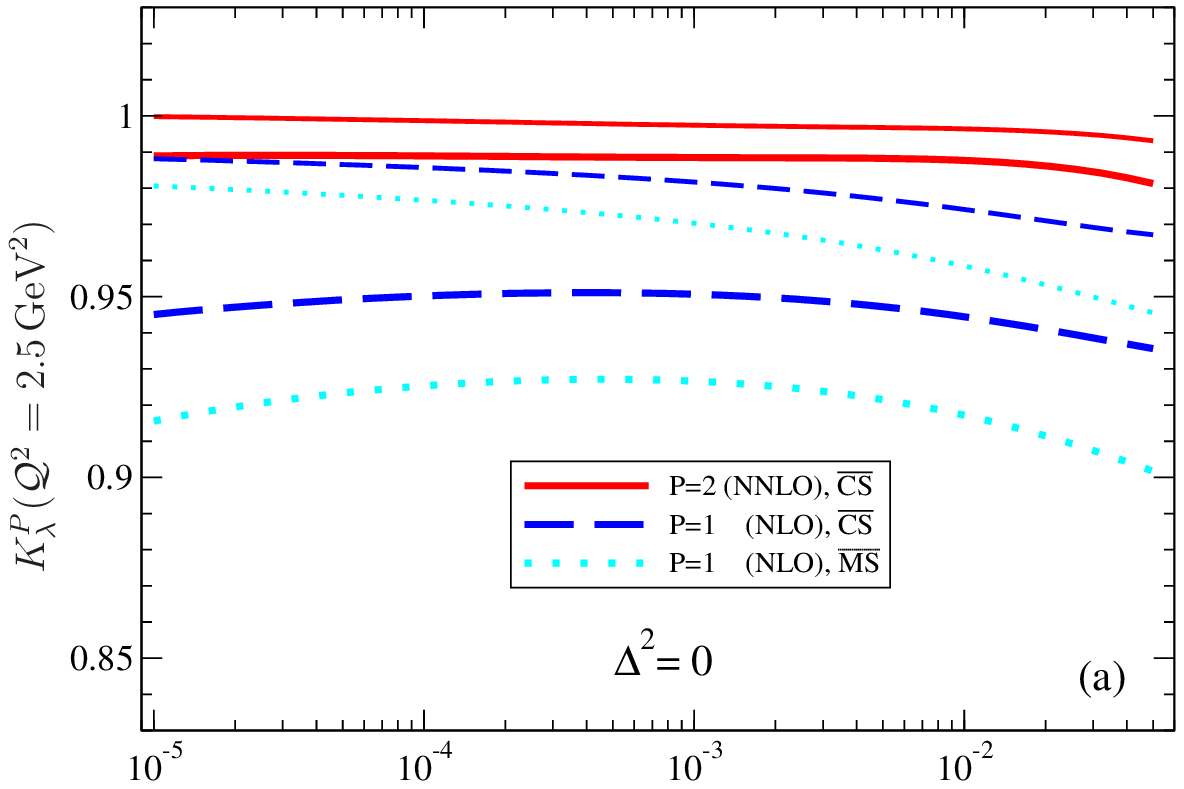}\hspace{2ex}\includegraphics[clip,scale=0.72]{fig1b.eps}
\includegraphics[clip,scale=0.72]{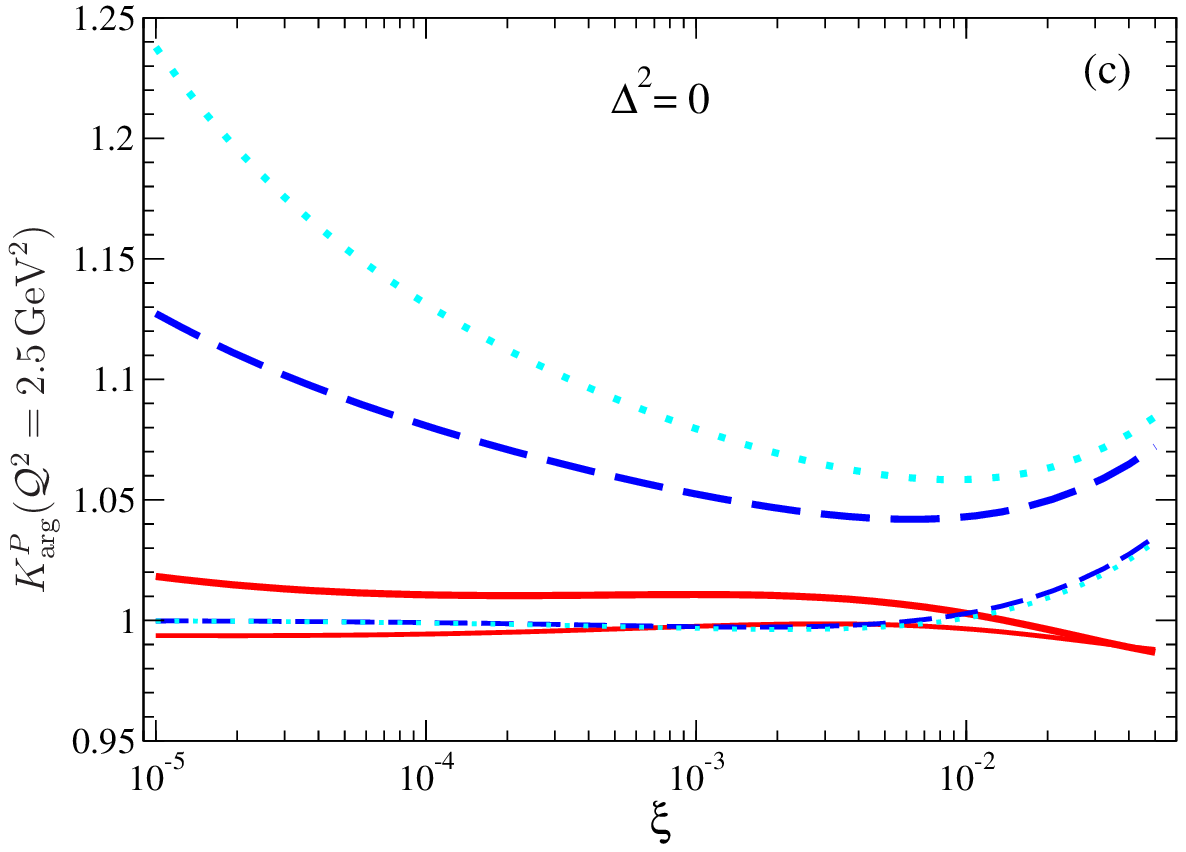}\hspace{2ex}\includegraphics[clip,scale=0.72]{fig1d.eps}
\caption{ \label{FigNNLO} The relative radiative corrections,
defined in Eq.\ (\ref{Def-Rrat}), are plotted versus $\xi$ for the
logarithm of the modulus [(a) and (b)] and phase [(c) and (d)] of
${^{\rm S}{\cal H}}$, see Eqs.\ (\ref{Res-ImReCFF}) and
(\ref{Ans-ConMom}), for $\Delta^2=0$ [(a) and (c)] and $\Delta^2=
-0.5\,\mbox{GeV}^2$  [(b) and (d)]: NNLO  (solid) as well as in
NLO for the $\overline{\rm CS}$ (dashed) and $\overline{\rm MS}$
(dotted) scheme. Thick (thin) lines refer to the ``hard''
(``soft'') gluon parameterization and we  always set $\mu= {\cal
Q}$ and  $\alpha_s({\cal Q}^2= 2.5\, \GeV^2)/\pi = 0.1$. }
\end{figure}
Usually, the GPDs are taken from some
non--perturbative (model) calculation or ansatz and plugged into
the CFFs at a given input scale. Since the scale and the
perturbative scheme are usually not specified, the matching of
perturbative and non--perturbative frameworks has its own
uncertainties. Let us first study the changes of the CFF
(\ref{Res-ImReCFF}) in a given scheme and input scale that appear
when one includes the next order. The changes to the modulus and
phase are appropriately measured by the $K$-factors:
\begin{eqnarray}
\label{Def-Rrat} K^P_{\lambda}=\frac{\ln \left|{^{\rm S}{\cal H}}^{{\rm N}^P{\rm
LO}}\right|}{\ln \left|{^{\rm S}{\cal H}}^{{\rm N}^{P-1}{\rm LO}}\right|}\,,
\qquad {K}^P_{\rm arg}= \frac{{\rm arg}\!\left(\! {^{\rm S}{\cal H}}^{{\rm
N}^P{\rm LO}}\!\right)}{{\rm arg}\!\left( {^{\rm S}{\cal H}}^{{\rm
N}^{P-1}{\rm LO}}\right)}\,.
\end{eqnarray}
Here ${^{\rm S}{\cal H}}^{{\rm N}^{P}{\rm LO}}$ denotes
the ${\rm N}^{P}{\rm LO}$ approximation, e.g., $P=0$ for ${\rm LO}$.
One should bear in mind that
$K$-factors actually measure the necessary reparameterization
of the GPD to fit the given experimental data.  We take the ansatz
(\ref{Ans-ConMom}) with ${^{\rm sea}\! F}(\Delta^2)={^{\rm G}\!F}(\Delta^2)$
and so the factorized $\Delta^2$ dependence essentially drops out in the
$K$--factors. The ratio of gluon GPD to quark one is controlled by the factor
$N_{\rm G}/N_{\rm sea}$ and, more importantly, by the differences of intercepts
$\alpha_{\rm G}(0)-\alpha_{\rm sea}(0)$. To study the influence of this ratio,
we distinguish two  cases:
\begin{eqnarray}
\mbox{H)\hspace{1cm} ``hard'' gluon:} && \hspace{1cm} N_G=0.4,\hspace{1cm} \alpha_{\rm G}(0) =\alpha_{\rm sea}(0)+0.1\, ,\\
\mbox{S)\hspace{1cm} \phantom{d}``soft'' gluon:}&& \hspace{1cm} N_G=0.3,\hspace{1cm} \alpha_{\rm G}(0) =\alpha_{\rm sea}(0)\, .
\end{eqnarray}
We will use these parameters and $\alpha_{\rm sea}(0)=1.1$ at
the input scale ${\cal Q}^2= 2.5\,\mbox{GeV}^2 $. Moreover, we set
$\mu={\cal Q }$ and independently of the considered approximation we
choose $\alpha_s(\mu_r^2= 2.5\,{\rm GeV}^2 ) = 0.1 \pi$ and set
the number of active flavors to three.

In Fig.\ \ref{FigNNLO} we depict for the typical kinematics of
HERA collider experiments, i.e., $10^{-5}\lesssim \xi \lesssim 5\cdot
10^{-2} $, the resulting $K$ factors for the logarithm of the modulus [(a) for
$\Delta^2=0$ and (b) for $\Delta^2=-0.5\,\mbox{GeV}^2$] and phase
[(c) for $\Delta^2=0$ and (d) for $\Delta^2=-0.5\,\mbox{GeV}^2$].
Here the thick and thin lines correspond to the ``hard'' and
``soft'' gluon parameterizations, respectively. We observe an almost flat $\xi$
dependence of the $K_\lambda$ factors in panels (a) and (b). This is not surprising,
since the essential contribution arises from the pomeron pole and the CFF behaves as:
\begin{eqnarray}
{^{\rm S}{\cal H}} \sim \left(\frac{1}{\xi}\right)^{\alpha(\Delta^2)}
\left[i + \tan\left(\frac{\pi}{2}(\alpha(\Delta^2)-1)\right)\right]
\quad\Rightarrow\quad
\ln |{^{\rm S}{\cal H}|} \cong \alpha(\Delta^2) \ln(1/\xi) + {\rm const.}\,.
\end{eqnarray}
For small $\xi$ this leads to the flatness we observe.
The size of perturbative NLO
corrections, see dashed ($\overline{\rm CS}$  scheme) and dotted
($\overline{\rm MS}$ scheme) lines, essentially depends on the
ratio of gluon to quark GPDs. Since the gluons are a new entry,
formally counted as NLO contribution, this finding is obvious and
goes along with the observation  that the perturbative corrections
strongly vary within the used parameterization of parton densities
in the Radyushkin GPD ansatz \cite{FreMcDStr02}. In the ``hard''
gluon parameterization the logarithm of the modulus reduces about
7--11\% [5--8\%] in the $\overline{\rm MS}$ [$\overline{\rm CS}$]
scheme, corresponding to the reduction of the modulus
itself in the range of 40--70\% [30--55\%],
where the drastic upper values correspond
to $\xi=10^{-5}$.

The relative radiative corrections to the phase
grow in the small $\xi$ region with decreasing $\xi$ and can be
of the order of up to 24\% [13\%] in the $\overline{\rm MS}$
[$\overline{\rm CS}$] scheme. These effects are related to the signs for NLO
Wilson coefficients, see Eqs.\ (\ref{Res-WilCoe-MS-NLO-V}),
(\ref{Res-WilCoe-MS-NLO-Vg}), (\ref{eq:NScV1}), and
(\ref{Def-Coe-NLO-G-V}). In the ``soft'' gluon parameterization the
NLO corrections are quite moderate for the modulus [(a) and (b)]
and negligible for the phase [(c) and (d)]. From all four panels
it can be realized that compared to $\overline{\rm MS}$ scheme in
the $\overline{\rm CS}$ one the NLO corrections are typically
reduced by  30--50\%. This reduction has been also observed in the
flavor non--singlet case \cite{Mue05a}. The NNLO corrections
(solid), compared to the NLO (dashed) ones, are drastically
reduced. For the ``soft'' gluon parameterization they are
practically negligible while for the ``hard'' gluon input they are
reduced to the 1--2\% level, except for the phase with
$\Delta^2=-0.5 {\rm GeV}^2$, where 5\% are reached at
$\xi=10^{-5}$.

Let us finally address the modification of the scale dependence
due to the higher order corrections.
We only consider here the $\overline{{\rm CS}}$ scheme and,
analogously  as in  Eq.\ (\ref{Def-Rrat}), we quantify the
relative changes due to the evolution by the ratios
\begin{eqnarray}\label{Def-Rrat-dot}
\dot{K}^P_{\lambda}= \frac{d \ln\left|{^{\rm S}{\cal H}}^{\rm {N}^P{\rm
LO}}\right|}{d\ln{\cal Q}^2} {\Bigg/}\frac{d \ln\left|{^{\rm S}{\cal
H}}^{\rm {N}^{P-1}{\rm LO}}\right|}{d\ln{\cal Q}^2}\,, \;
\dot{K}^P_{\rm arg}=
  \frac{d\, {\rm arg}\left({^{\rm S} {\cal H}}^{\rm {N}^P{\rm LO}}\right)}{d\ln{\cal Q}^2}
  \Bigg/
  \frac{d\, {\rm arg}\left( {^{\rm S} {\cal H}}^{\rm {N}^{P-1}{\rm LO}}\right)}{d\ln{\cal
  Q}^2}.
\end{eqnarray}
For the (exact) evolution of $\alpha_s({\cal Q})$ we take the same scale setting and
initial condition  as above. However, the conformal moments (\ref{Ans-ConMom}) are evolved in the
$\overline{\rm CS}$ scheme, starting at the input scale  ${\cal Q}_0^2= 1\,
{\rm GeV}^2$, to ${\cal Q}^2= 4\, {\rm GeV}^2$. The non--leading logs in the solution of
the evolution equation (\ref{Def-RGE-1}) are expanded with respect
to $\alpha_s$ and are consistently combined with the
Wilson--coefficients (\ref{Res-WilCoe-Exp-CS-SI}) in the considered
order.  The unknown NNLO mixing term
$\Delta_{jk}^{\overline{{\rm CS}}}$ in Eq.\ (\ref{Def-RGE-1}) is
neglected. This mixing can be suppressed at the input scale by
an appropriate initial condition  and so we expect only a minor
numerical effect; see also Ref.\ \cite{Mue98}.
\begin{figure}[t]
\includegraphics[clip,scale=0.72]{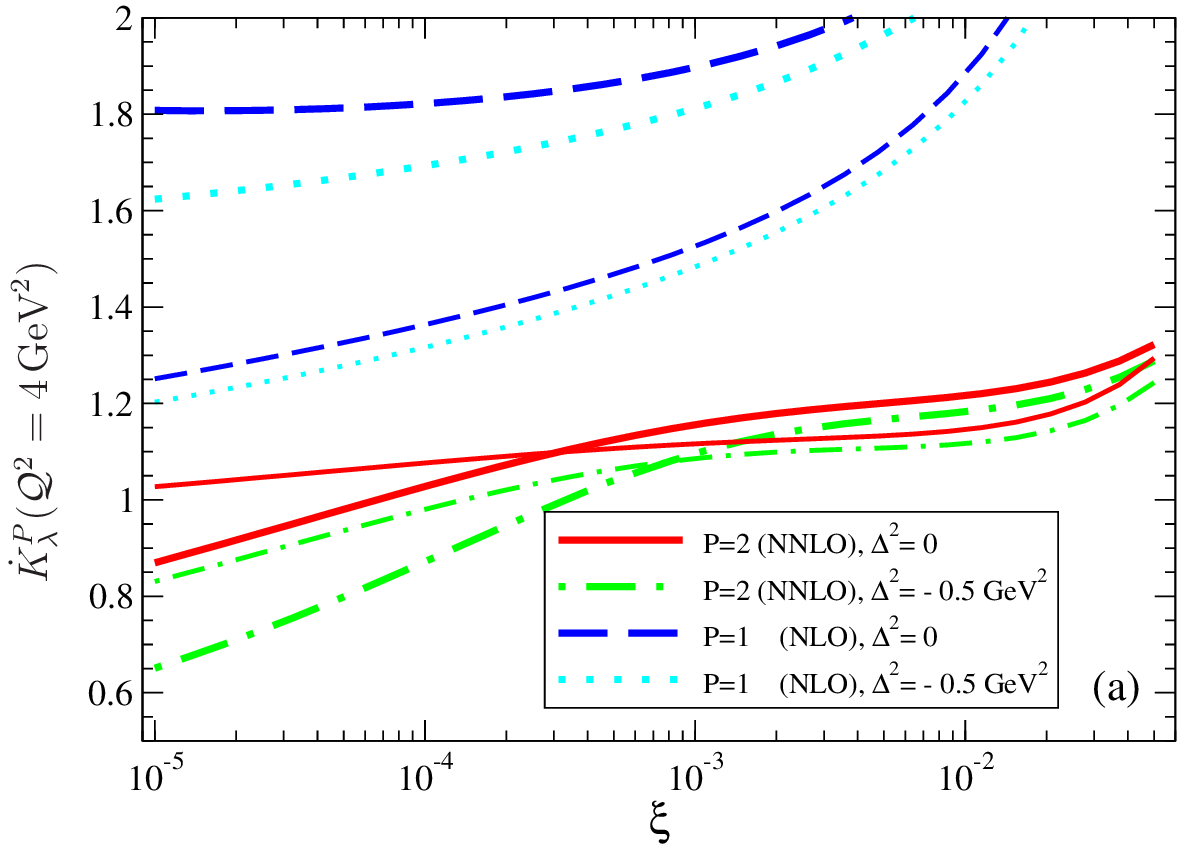}
\includegraphics[clip,scale=0.72]{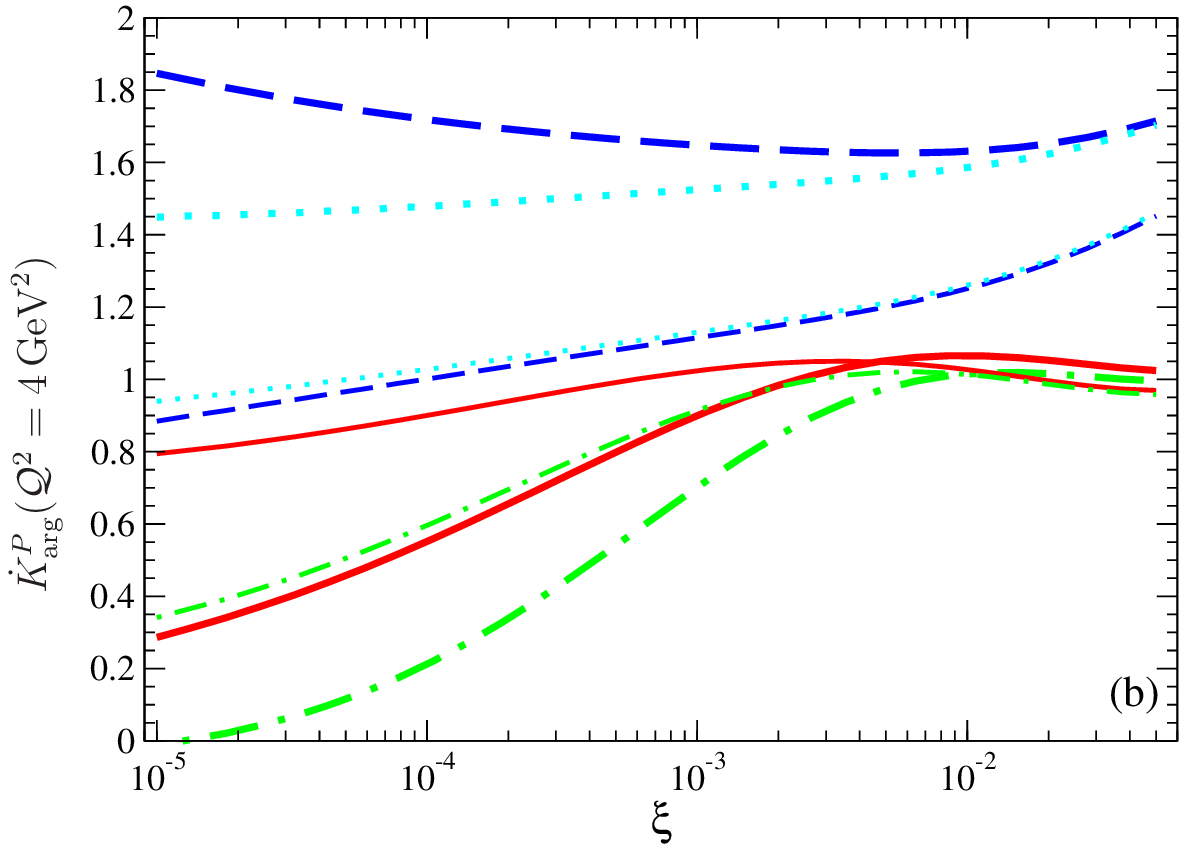}
\caption{\label{Fig-ScaDep} The relative change of scale
dependence, cf.\ Eq.\ (\ref{Def-Rrat-dot}), in the $\overline{\rm
CS}$ scheme at NLO (dashed, dotted)  and NNLO (solid, dash--dotted)
versus $\xi$ is depicted for the logarithm of the modulus (a) and phase (b) of the
CFF (\ref{Res-ImReCFF}) with $\Delta^2=0$ (dashed, solid)  and
$\Delta^2=-0.5\, {\rm GeV}^2$ (dotted, dash--dotted) and ${\cal Q}^2 =4\, \GeV^2$.
We set $\mu = {\cal Q}$, $\alpha_s(\mu_r^2= 2.5\, \GeV^2 ) /\pi =
0.1$ and took the input (\ref{Ans-ConMom}) at the scale ${\cal
Q}^2_{0} = 1\, \GeV^2 $. Thick and thin lines correspond again to ``hard'' and
``soft'' gluonic input.}
\end{figure}
The dashed and dotted lines in Fig.\ \ref{Fig-ScaDep} show that in
NLO the scale dependence changes can be rather large even  of
about 100\% or more. In general the relative radiative corrections
to NNLO are getting smaller. For instance, the NNLO corrections in
panel (a) are almost negligible for the ``soft'' gluonic input
with $\Delta^2=0$ (thin solid), they increase, however, for
$\Delta^2=-0.5\, {\rm GeV}^2$ and are becoming large for the
``hard''  gluonic input, e.g., about $-35\%$ at $\xi=10^{-5}$
(thick dash-dotted). Note that these large corrections at very
small $\xi$ are essentially caused by those of the anomalous
dimensions in the vicinity of $j=0$, corresponding to the large
corrections of
the gluon splitting kernels at small $x$, reported in
\cite{VogMocVer04}. The same sources also cause the huge NNLO
corrections to the phase in panel (b). We remark that the modulus
of $^{\rm S}{\cal H}$ is dominated by its imaginary part for which radiative
corrections are milder than for the real part. The real part and so
also the phase at very small $\xi$ are rather strongly affected by
the NNLO corrections to anomalous dimensions. On the other hand
for $5\cdot 10^{-4} \lesssim \xi$ and $5\cdot 10^{-3} \lesssim
\xi$ the radiative NNLO corrections to the logarithm of the
modulus (a) and phase (b), respectively, are rather mild  (solid
and dash--dotted lines). Restricted to these kinematics our findings support
the convergence of the perturbative series.

\section{Summary}
\label{Sec-Con}

In this letter we have studied NLO and NNLO corrections to deeply
virtual Compton scattering in the small $\xi$ region. We confirmed
that large radiative corrections at NLO can appear, reported
before, and clarified their source which is entirely tied to the
gluonic sector. In particular, if the gluon distribution starts to have a
steeper increase at small $\xi$ than the quark ones, the NLO
corrections will be dominated by the  negative NLO gluon
contribution and so the modulus of $^{\rm S}{\cal H}$ will drastically
reduce. On the other hand, if the gluon contribution is relatively
small, already the NLO corrections are moderate. In any case the
NNLO corrections are becoming moderate or even small at a given
input scale, even at a few $\mbox{GeV}^2$. This fact supports the
perturbative framework of DVCS.

The situation with respect to the scale dependence is not so
conclusive. Going from LO to NLO we observe in general a big
enhancement that arises from the large corrections to the
anomalous dimensions, cf.\ \cite{VogMocVer04}. To NNLO they will
be reduced and the relative changes for the logarithm of the
modulus are getting reasonable but grows to be large with decreasing
$\xi$. Note that in this region the NNLO gluonic evolution effects
are comparable in size with the NLO ones \cite{VogMocVer04}.
Also the NLO radiative corrections to the scale dependence of the phase of
$^{\rm S}{\cal H}$  are rather large, in
particular for $\Delta^2=0$,  at the scale of 4 $\mbox{GeV}^2$. To
NNLO accuracy  the convergency  improves for
$5\cdot 10^{-3} \lesssim \xi$. Unfortunately, at smaller values of $\xi$ the
convergency is lost.  These large corrections due to evolution at
small $\xi$ are certainly related to those found in DIS \cite{VogMocVer04}.

If one is interested to access GPDs from the DVCS cross section
measurement at small $\xi$, only the modulus of $^{\rm S}{\cal H}$ is
essential. In that case perturbation theory seems to work in the sense that
NNLO corrections of the Wilson--coefficients are negligible. They are,
however, important for the scale violating effects for $\xi
\lesssim 5\cdot 10^{-4}$ (at relatively low ${\cal Q}^2
\sim 4\, \mbox{GeV}^2$). We also conclude that the photon and
vector--meson leptoproduction data taken by the  H1 and ZEUS
collaborations should be perturbatively analyzed at NLO
\cite{IvaSzyKra04}. To our best knowledge a common perturbative
analysis  has not been done so far. To achieve this in a simple and
numerical stable manner, the Mellin--Barnes integral
representation seems to be preferred.

This project has been supported by  the U.S. National Science
Foundation under grant no. PHY--0456520, German Research Foundation (DFG), and
Croatian Ministry  of Science, Education and Sport under the contract no.
0119261, as well as The National Foundation for Science, Higher Education and
Technological Development of the Republic of Croatia under the contract 01.03./02.



\end{document}